\documentclass[article]{IEEEtran}
\usepackage{xcolor}
\usepackage{cite}
\usepackage{graphicx}
\usepackage{caption} 
\usepackage{subcaption} 
\usepackage[english]{babel} 
\usepackage{amsmath,amsfonts,amsthm,bm}
\usepackage{wrapfig}
\usepackage{multirow}
\usepackage[framed,numbered,autolinebreaks,useliterate]{mcode}
\usepackage{mwe}
\usepackage{comment}
\usepackage{url}
\usepackage{cite}

\ifCLASSINFOpdf
 
\else
  
\fi

\begin{document}
\title{A Compact Model of Threshold Switching Devices for Efficient Circuit Simulations}
\author{
\IEEEauthorblockN{ M. M. Al Chawa$^{*}$,~\IEEEmembership{Member,~IEEE}, and A. S. Demirko,~\IEEEmembership{Member,~IEEE}
}
\\
}

\IEEEoverridecommandlockouts

\author{Mohamad Moner~Al Chawa,~\IEEEmembership{Member,~IEEE,}  ~Ahmet~S.~Demirkol,~\IEEEmembership{Senior~Member,~IEEE}
              and~Ronald~Tetzlaff,~\IEEEmembership{Senior~Member,~IEEE}
\IEEEcompsocitemizethanks{\IEEEcompsocthanksitem M. M. Al Chawa, and R.Tetzlaff are with the Institute of Circuits and Systems, Technische Universität Dresden, Dresden, 01069 .\protect\\
E-mail:mohamad\_moner.al\_chawa@tu-dresden.de
}
}

\IEEEoverridecommandlockouts

\author{Mohamad Moner~Al Chawa,~\IEEEmembership{Member,~IEEE,}
        Daniel~Bedau,
        Ahmet~S.~Demirkol, 
        James~W.~Reiner,
        Derek~A.~Stewart,~\IEEEmembership{Senior~Member,~IEEE,}
        Michael~K.~Grobis,
        and~Ronald~Tetzlaff,~\IEEEmembership{Senior~Member,~IEEE,}

 \IEEEcompsocitemizethanks{\IEEEcompsocthanksitem M. M. Al Chawa, A. S. Demirkol and R. Tetzlaff are with the Institute of Circuits and Systems, Technische Universität Dresden, Dresden, 01069, Germany. \protect
 \\
 E-mail:mohamad\_moner.al\_chawa@tu-dresden.de  }
\IEEEcompsocitemizethanks{\IEEEcompsocthanksitem D. Bedau, J. W. Reiner, D. A. Stewart and M. K. Grobis are with Western Digital San Jose Research Center, CA, USA. \protect
\\
E-mail:daniel.bedau@wdc.com }

}

\maketitle

\begin{abstract}
In this paper, we present a new compact model of threshold switching devices which is suitable for efficient circuit-level simulations. First, a macro model, based on a compact transistor based circuit, was implemented in LTSPICE. Then, a descriptive model was extracted and implemented in MATLAB, which is based on the macro model. This macro model was extended to develop a physical model that describes the processes that occur during the threshold switching. The physical model derived consider a delay structure with few electrical components near to the second junction.
The delay model incorporates an internal state variable, which is crucial to transform the descriptive model into a compact model and to parameterize it in terms of electrical parameters that represent the component's behavior. Finally, we applied our model by fitting $i\text{--}v$ measured data of an OTS device manufactured by Western Digital Research.
\end{abstract}
\section{Introduction}
The ovonic threshold switch (OTS) is a promising two-terminal nanodevice based on chalcogenide alloys.
OTS devices typically demonstrate a current controlled negative differential resistance (NDR) characteristic on their DC $i\text{--}v$ loci, and therefore, they can exhibit very sharp and fast transition between on and off states that makes them attractive for many applications, including as selectors for memory cells, as fast switches, or as devices for neuromorphic computing\cite{tuma2016stochastic, song2019ovonic, readEvaluatingOvonicThreshold2021}.  
\begin{figure}[b!]
       \includegraphics[width=1.0\columnwidth]{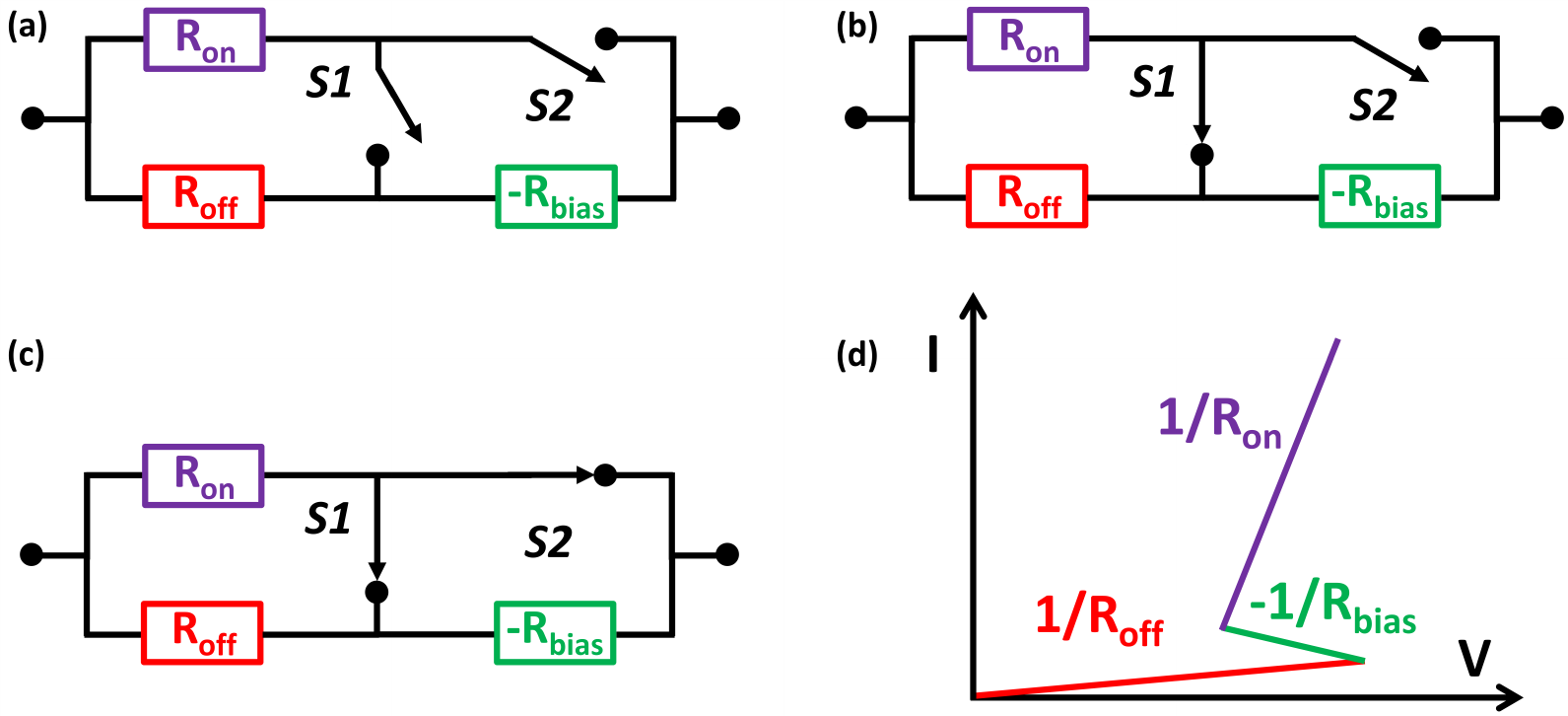}
    \caption{ Conceptual block diagram of the threshold switching device: $R_{\text off} >> R_{bias} >> R_{on}$. (a) Off state ($S_1=off$ and $S_2=off$); (b) Snapback state ($S_1=on$ and $S_2=off$); (c) On state ($S_1=on$ and $S_2=on$); (d) $i\text{--}v$ for (a-c).}
        \label{fig:block}
\end{figure}
\begin{figure}[t!]
    \includegraphics[width=0.9\columnwidth]{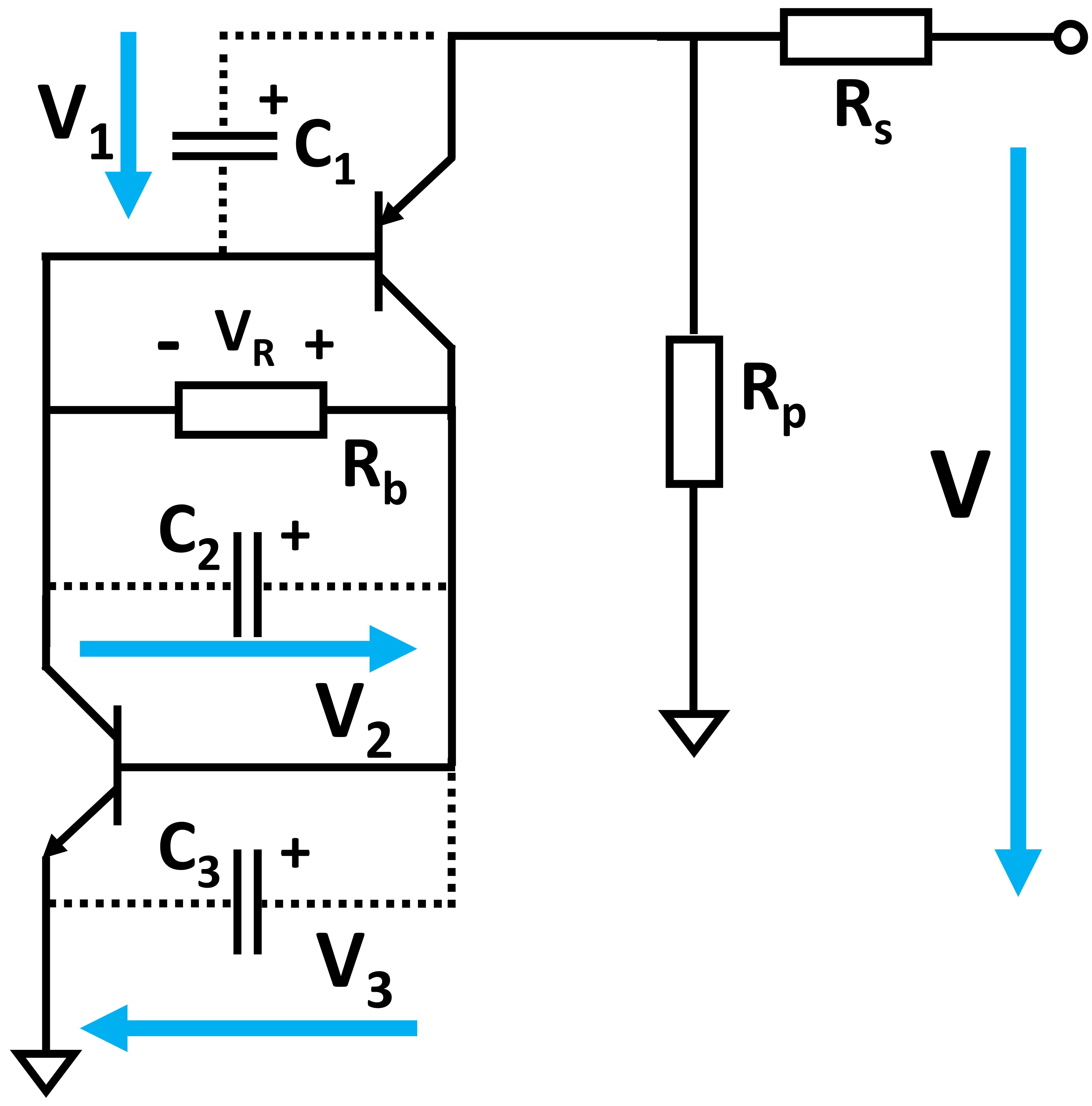}
    \caption{Threshold switching (OTS) device equivalent circuit ($R_{S}=0$ and $R_{\text P}= \infty$ for ideal device).}
    \label{fig:circuit}
\end{figure}
\begin{figure*}[t!]
     \centering
      \begin{subfigure}[t]{0.45\textwidth}
         \centering
              \includegraphics[width=\textwidth]{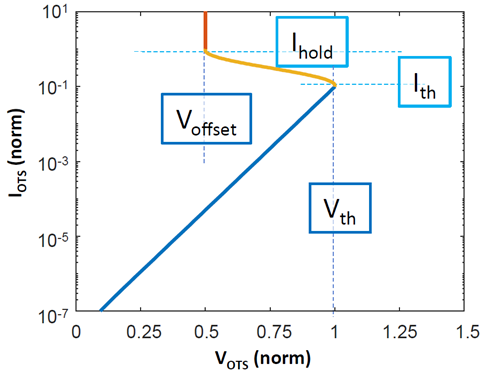}
         \caption{}
         \label{fig:a11}
     \end{subfigure}
    \hfill
     \begin{subfigure}[t]{0.48\textwidth}
         \centering
                \includegraphics[width=\textwidth]{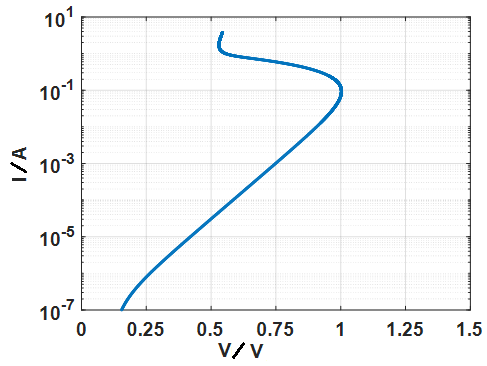}
         \caption{}
         \label{fig:b11}
     \end{subfigure}
     \hfill
     \begin{subfigure}[t]{0.45\textwidth}
         \centering
               \includegraphics[width=\textwidth]{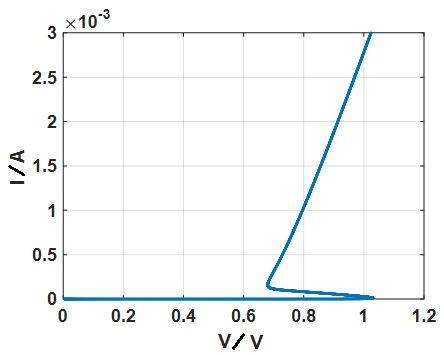}
        \caption{}
         \label{fig:c11}
     \end{subfigure}
       \hfill
     \begin{subfigure}[t]{0.41\textwidth}
         \centering
                   \includegraphics[width=\textwidth]{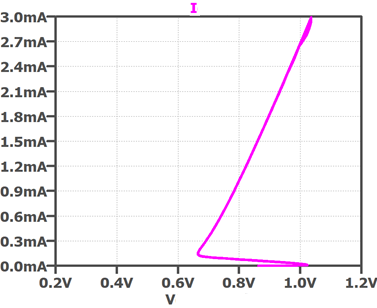}                    \caption{}
         \label{fig:d11}
     \end{subfigure}
        \caption{(a) Definition of parameters on the $i\text{--}v$ curve of a typical OTS device. (b-c) Implemented model in MATLAB extracted from the macro model, (b) is plotted using logarithmic scale; (d) Macro model implemented in LTSPICE; the $i\text{--}v$ curve has been obtained using a current sweep ($R_{b}=5 k\Omega$,  $R_{p}=100 k\Omega$, and  $R_{s}=200 \Omega$).}
        \label{fig:block221}
\end{figure*}
\begin{figure*}[t!]
     \centering
      \begin{subfigure}[t]{0.48\textwidth}
         \centering
              \includegraphics[width=\textwidth]{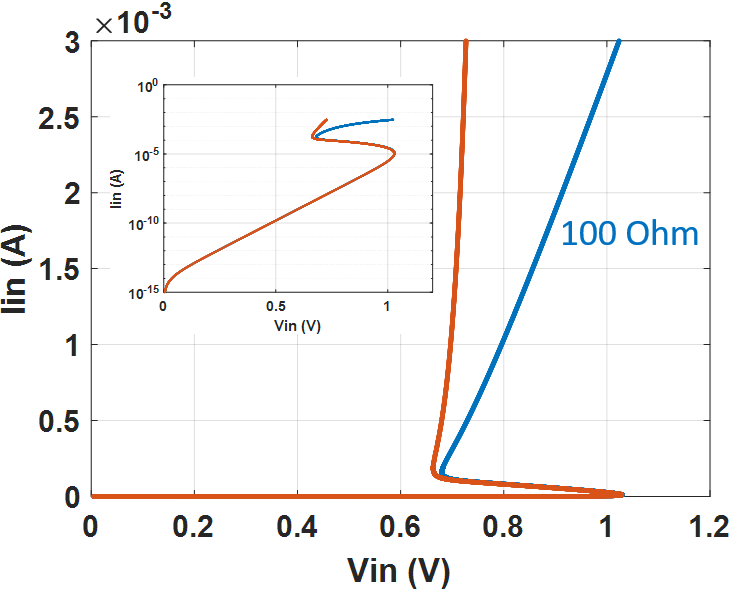}
         \caption{}
         \label{fig:parameters1}
     \end{subfigure}
    \hfill
     \begin{subfigure}[t]{0.48\textwidth}
         \centering
                \includegraphics[width=\textwidth]{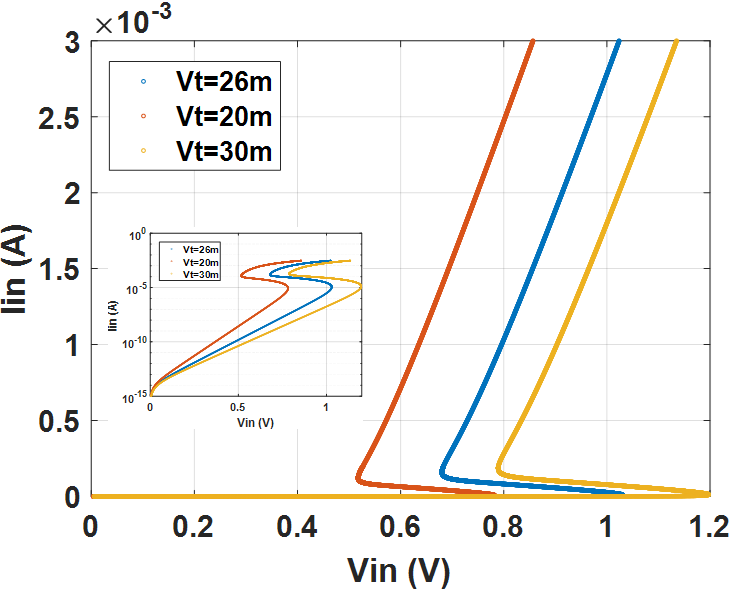}
         \caption{}
         \label{fig:parameters2}
     \end{subfigure}
     \hfill
     \begin{subfigure}[t]{0.48\textwidth}
         \centering
               \includegraphics[width=\textwidth]{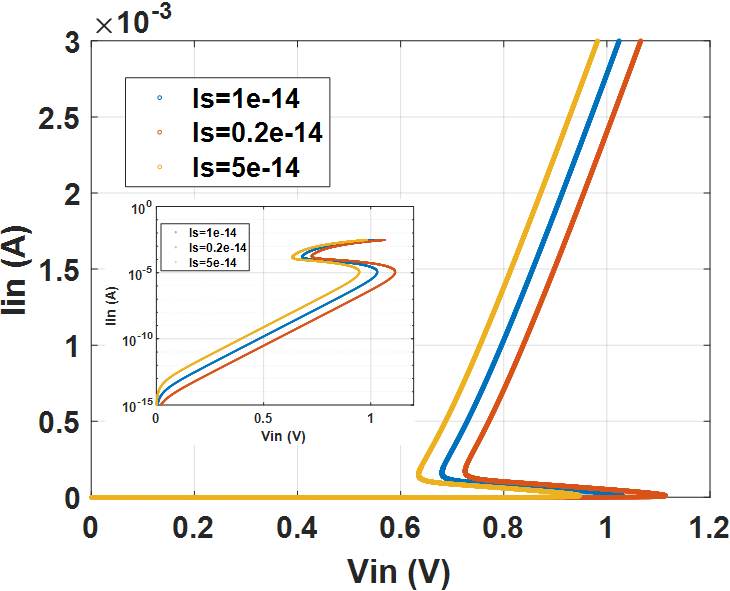}
        \caption{}
         \label{fig:parameters3}
     \end{subfigure}
       \hfill
     \begin{subfigure}[t]{0.48\textwidth}
         \centering
                   \includegraphics[width=\textwidth]{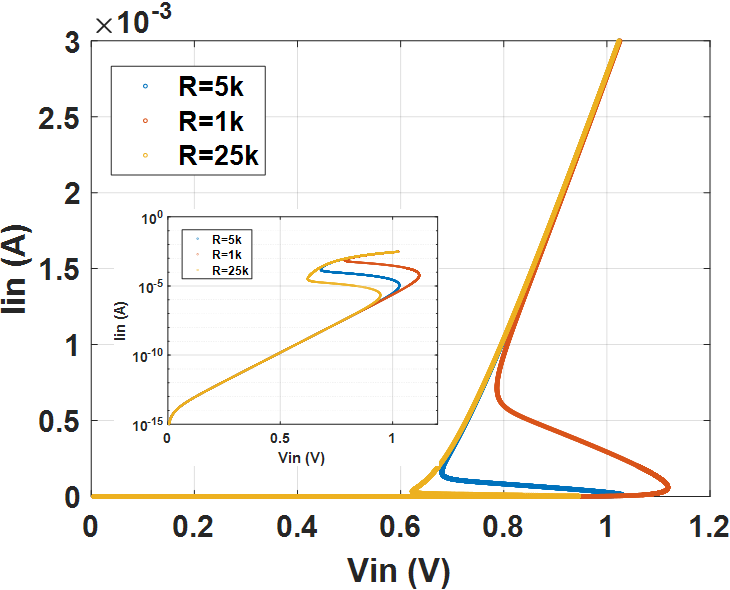}
         \caption{}
         \label{fig:parameters4}
     \end{subfigure}
        \caption{Model Parameter Variation. (a) Adding a resistor in series, $R_s=100 \Omega$; (b) Thermal voltage; (c) Saturation current; (d) Snapback resistor. The insets show the results using logarithmic scale for the current.}
        \label{fig:parameters}
\end{figure*}
\begin{figure}[tb!]
    \includegraphics[width=1.0\columnwidth, height=4cm]{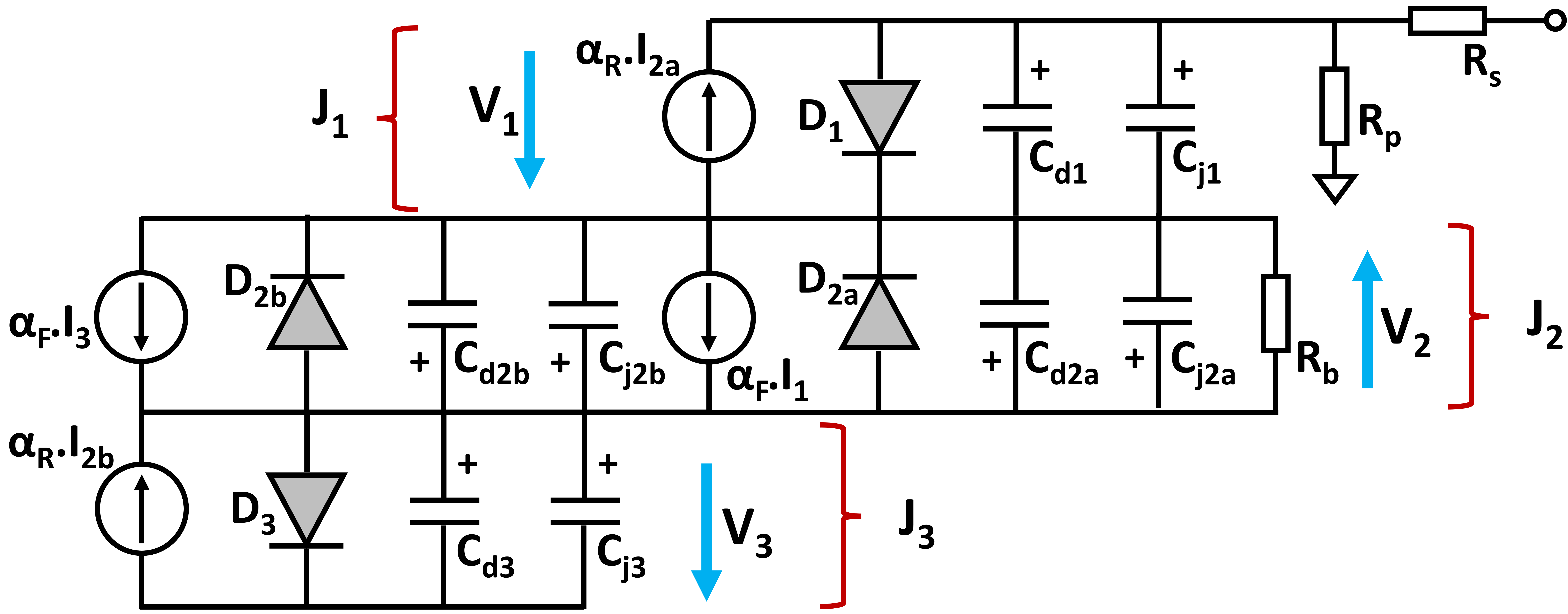}
    \caption{OTS device cell circuit.}
    \label{fig:newcircuit}
\end{figure}
 \begin{figure*}
     \centering
      \begin{subfigure}[b]{0.49\textwidth}
         \centering
                          \includegraphics[width=\textwidth]{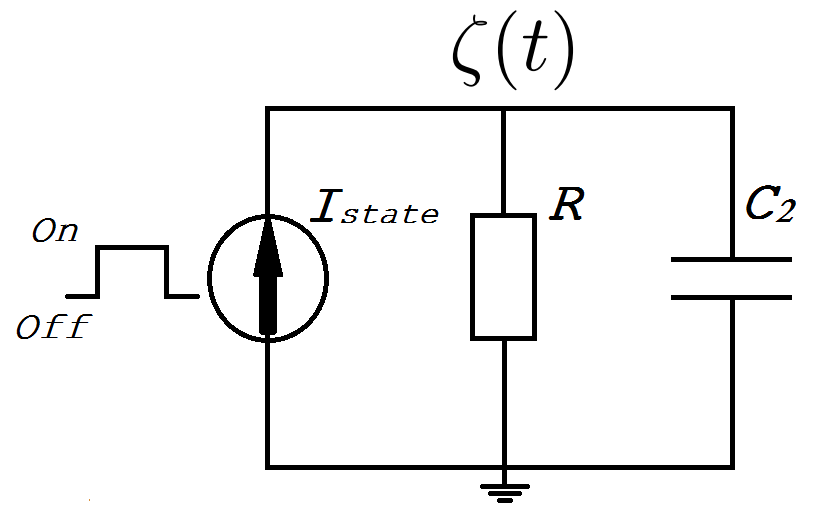}
         \caption{}
         \label{fig:a1}
     \end{subfigure}
     \hfill
     \begin{subfigure}[b]{0.5\textwidth}
         \centering
                    \includegraphics[width=\textwidth]{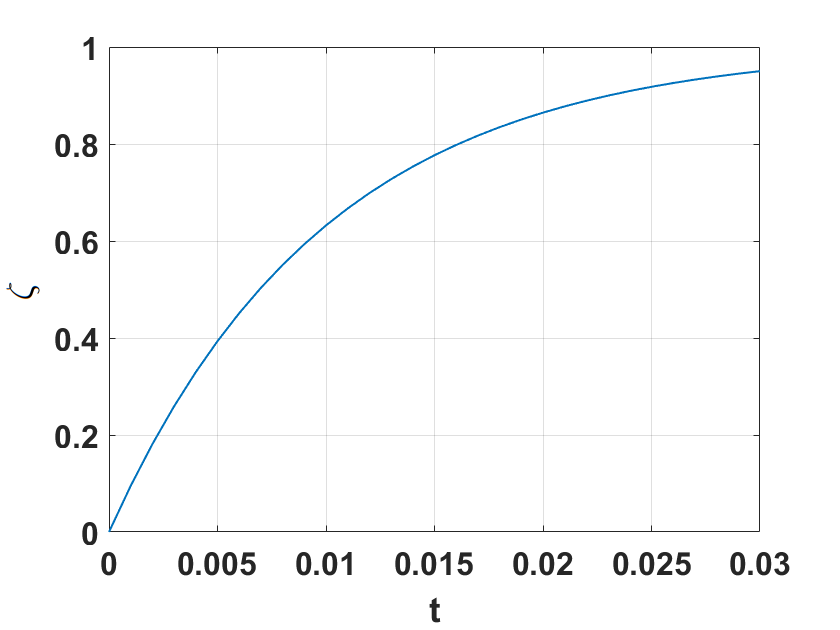}
         \caption{}
         \label{fig:state}
     \end{subfigure}
            \caption{
       Delay Model. (a) Internal state model extracted from $J_2$ in Fig. \ref{fig:newcircuit}: $C_{2}=1nF$, $R_b=1M\Omega$ and $I_{state}=1\mu A$; (b) variable state $\zeta (t)$ (which is treated as a voltage by the simulator) vs. time (a.u.). }
        \label{fig:delay model}
\end{figure*}
\begin{figure*}
     \centering
      \begin{subfigure}[b]{0.496\textwidth}
         \centering
                  \includegraphics[width=\textwidth]{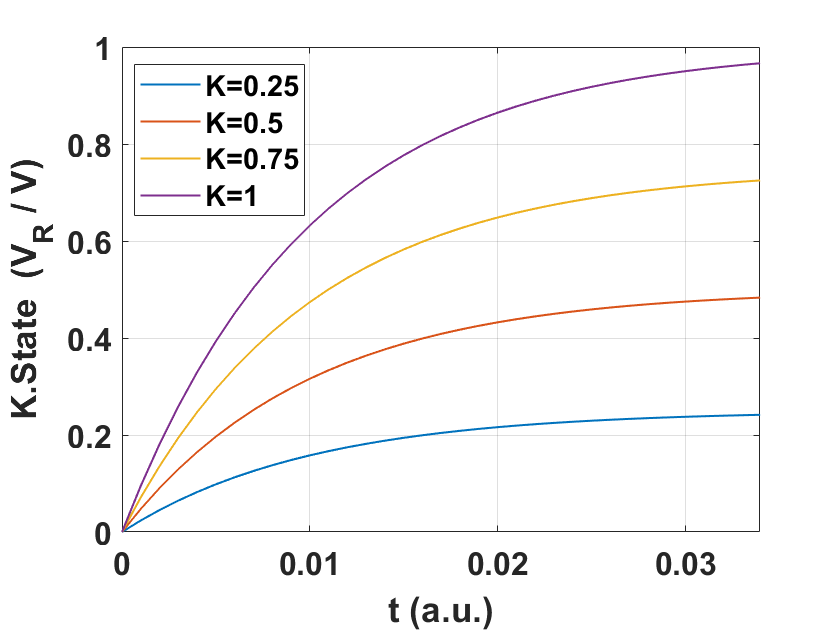}
         \caption{}
         \label{fig:Kless1}
     \end{subfigure}
     \hfill
     \begin{subfigure}[b]{0.496\textwidth}
         \centering
                    \includegraphics[width=\textwidth]{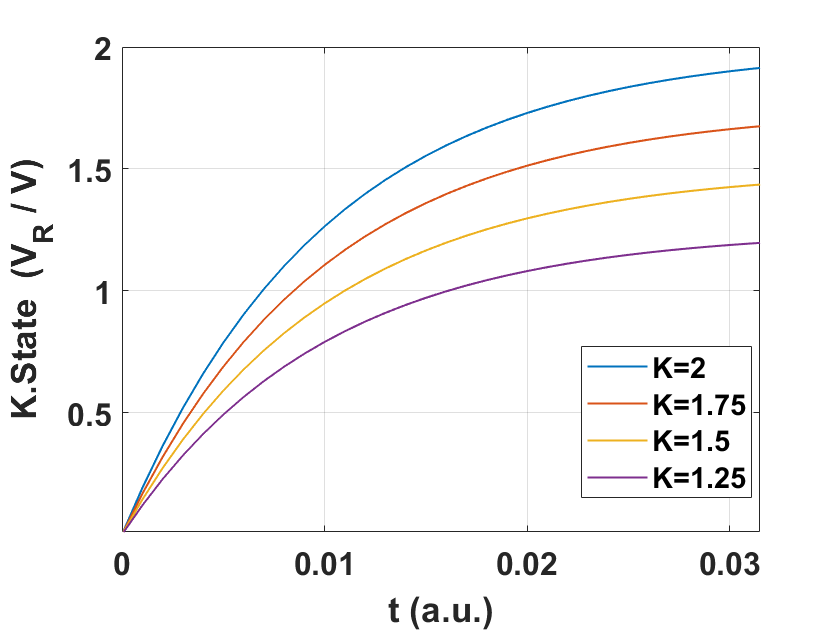}
         \caption{}
         \label{fig:Kmore1}
     \end{subfigure}
            \caption{
       Internal State Variable $v_{R}=K\cdot \zeta$ (a) $K<1$ 
 ; (b) $K>1$ }
        \label{fig: Internal State Variable}
\end{figure*}
\begin{figure*}[t!]
     \centering
      \begin{subfigure}[t]{0.48\textwidth}
         \centering
              \includegraphics[width=\textwidth]{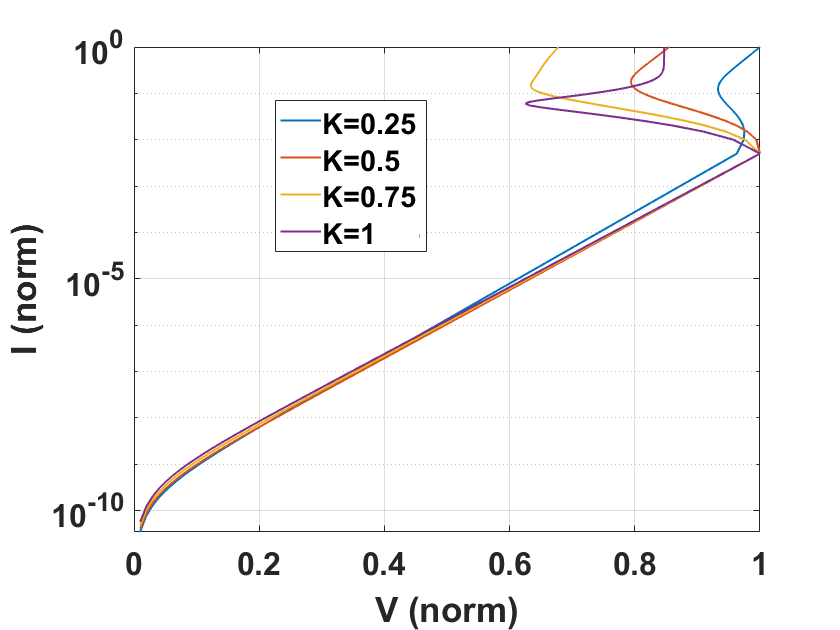}
         \caption{}
         \label{fig:parameters11}
     \end{subfigure}
    \hfill
     \begin{subfigure}[t]{0.48\textwidth}
         \centering
                \includegraphics[width=\textwidth]{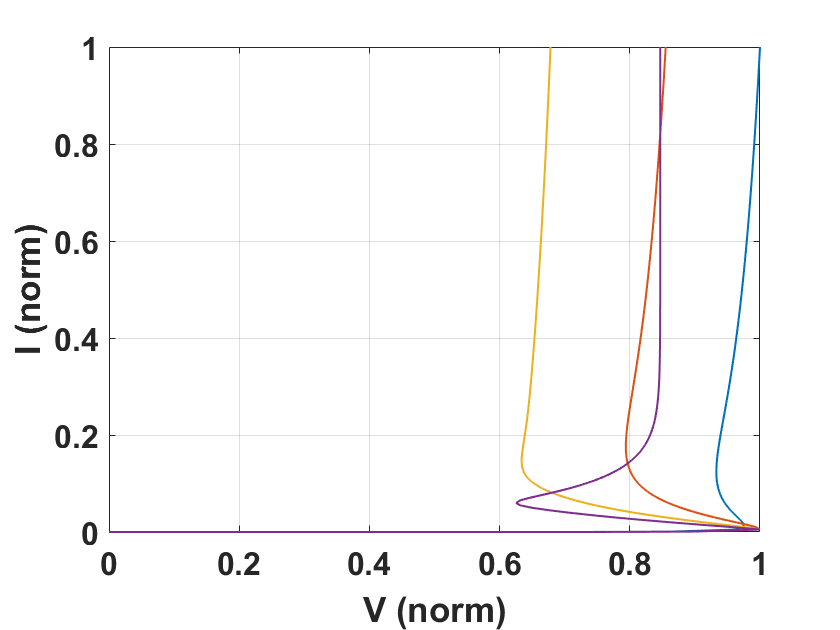}
         \caption{}
         \label{fig:parameters21}
     \end{subfigure}
     \hfill
     \begin{subfigure}[t]{0.48\textwidth}
         \centering
               \includegraphics[width=\textwidth]{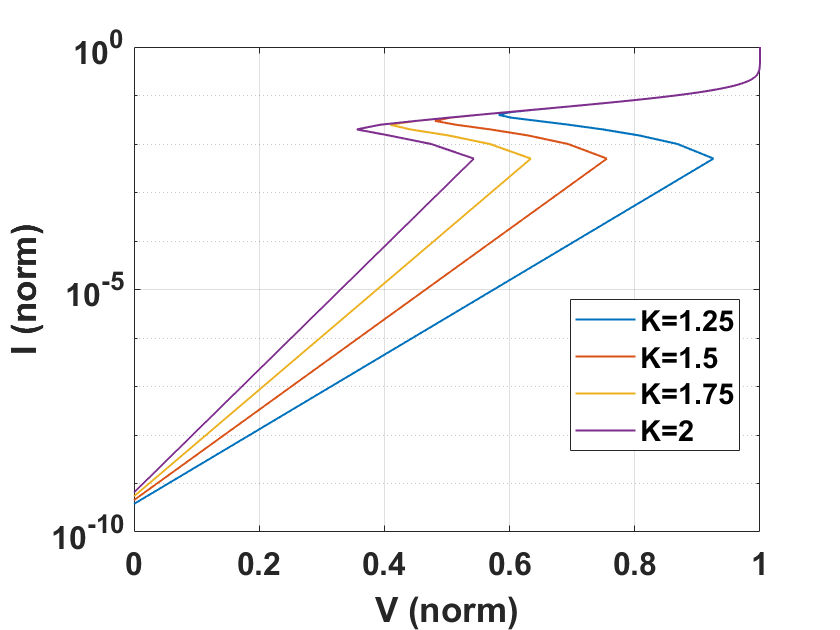}
        \caption{}
         \label{fig:parameters31}
     \end{subfigure}
       \hfill
     \begin{subfigure}[t]{0.48\textwidth}
         \centering
                   \includegraphics[width=\textwidth]{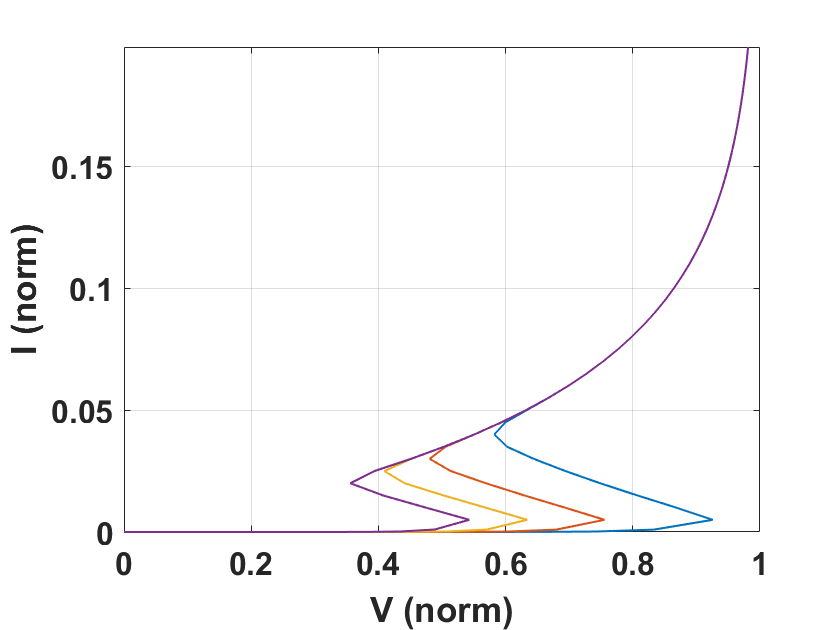}
         \caption{}
         \label{fig:parameters41}
     \end{subfigure}
        \caption{
       $i\text{--}v$ characteristic curves with different sets of $K$ using \eqref{eq:final model}.
       (a-b) $K<1$; (c-d) $K>1$.}
        \label{fig:i-v simulation}
\end{figure*}
\begin{figure*}
     \centering
      \begin{subfigure}[b]{0.496\textwidth}
         \centering
                  \includegraphics[width=\textwidth]{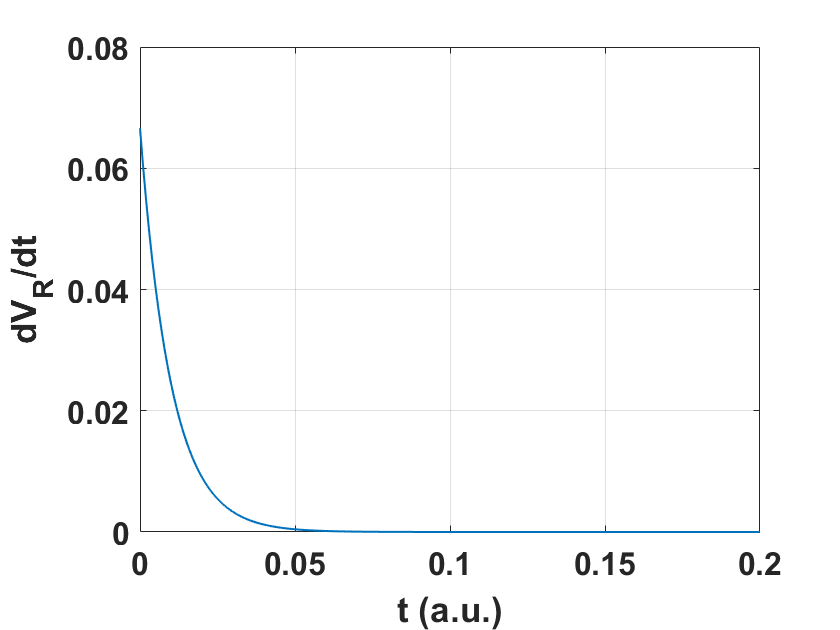}
         \caption{}
         \label{fig:dv}
     \end{subfigure}
     \hfill
     \begin{subfigure}[b]{0.496\textwidth}
         \centering
                    \includegraphics[width=\textwidth]{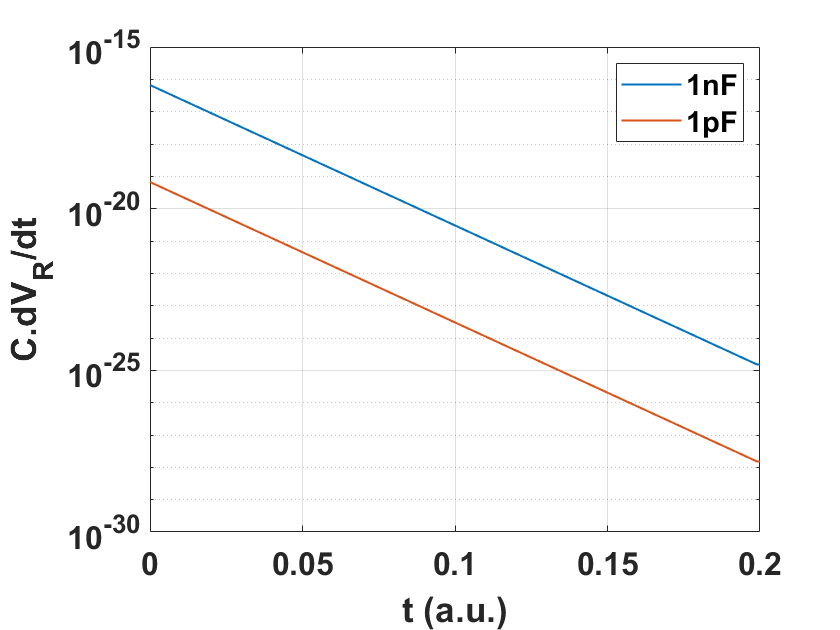}
         \caption{}
         \label{fig:charge model}
     \end{subfigure}
            \caption{
        (a) Derivative of the scaled internal state variable ($K=0.7$); (b) Charge Model for two values of capacitance ($1nF$ and $1pF$).}
        \label{fig:Q}
\end{figure*}
\begin{figure*}
     \centering
      \begin{subfigure}[b]{0.496\textwidth}
         \centering
                  \includegraphics[width=\textwidth]{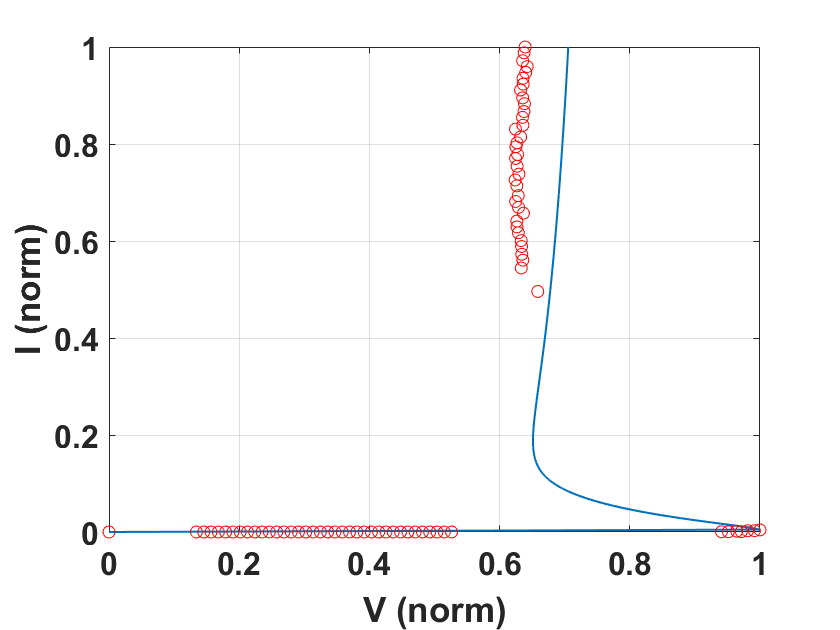}
         \caption{}
         \label{fig:Kless2}
     \end{subfigure}
     \hfill
     \begin{subfigure}[b]{0.496\textwidth}
         \centering
                    \includegraphics[width=\textwidth]{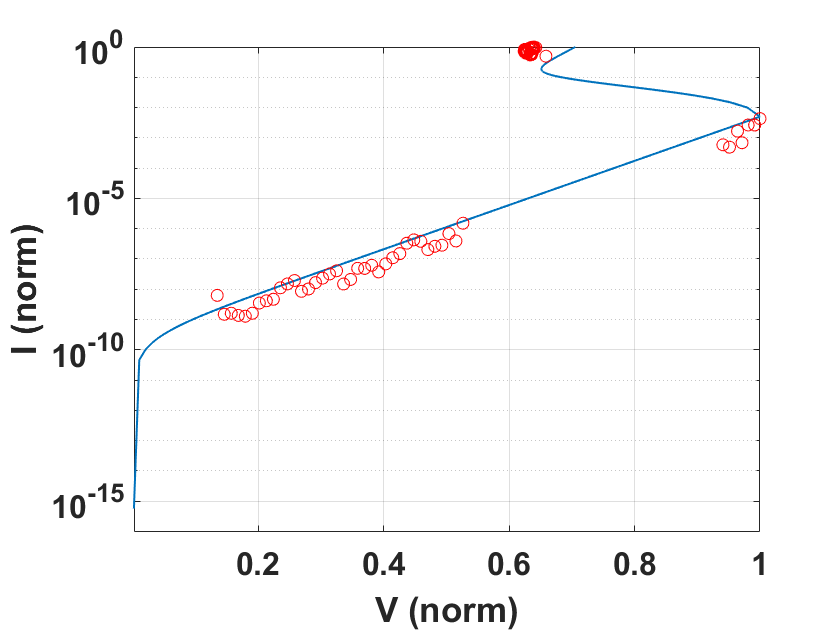}
         \caption{}
         \label{fig:Kmore2}
     \end{subfigure}
            \caption{
       $i\text{--}v$ curve, red dots are experimental measurements (symbols) and modeled using \eqref{eq:final model} (blue line) for an OTS device fabricated by Western Digital Research.
       }
        \label{fig:fitting}
\end{figure*}
While threshold switching in chalcogenide glasses has been known for a long time, the exact mechanism is still unknown. Although several models capture the phenomenological behavior of OTS cells very well \cite{zhu_mrs_2019}, they often tend to be not very suitable for practical circuit design applications which need an accurate, but fast, model of the device expressed in a way that naturally fits into an analog simulation pipeline. There is a trade-off between the physical complexity that describes the device and the accuracy of the model representing it. The ability to implement complicated models in circuit simulators represents a challenge. These challenges have been historically addressed in modelling by using the classical current-voltage description \cite{9642394, 9629238}, or Chua's original proposal in flux and charge \cite{8357574,8713508, 9478939, 9181155, al2018exploring}. Both descriptions are shown to be equivalent \cite{corinto2015theoretical}, and, for the sake of usual convention, we will use in this paper the current and voltage. 
Typical approaches use transport models\cite{9405114,9366291} that model the charge carrier dynamics using differential equations to simulate the behavior of an OTS device. These physics based models are complex and do not run natively in circuit simulators, and are not well suited as compact models for circuit simulations. An example of  a non-physical model of PCM for low-computational cost, as an alternative model, has been demonstrated in \cite{9031725}.
However, in this contribution, a 2T-1R macromodel for threshold switching devices, such as an OTS, has been implemented in LTSPICE. In addition, a mathematical description of the macromodel was made, including the design of a new circuit based on the 2T-1R circuit. Later, an internal state variable was extracted to convert the descriptive model into a compact model for threshold switching devices.
The proposed model is an accurate tool for circuit designers since It has been  implemented in a circuit simulator, LTSPICE. For example, OTS devices can be used as selectors  for memory cell
cross points.
Finally, the compact model was applied to fit $i\text{--}v$ measurement data obtained from an OTS device manufactured by Western Digital Research.
\section{Model Description and Implementation}
\label{sec:model}
The threshold switching device has three states: off, on, and snapback\cite{adler1980threshold, zalden2016picosecond}. Let us consider a resistor for each state, $R_\text{off}$, $R_\text{on}$, $-R_\text{bias}$, where the values of these resistors follows according to
\begin{equation}
    R_\text{off} >> R_\text{bias} >> R_\text{on} ~ .
\end{equation}
Fig. \ref{fig:block} shows a conceptual block diagram of these three states. In the ideal case, we have $R_\text{off}=\infty$ and $R_\text{on}=0$ which would result in slope values $0$ and $\infty$ in the off and the on states, respectively. 
For the Off state, both switches are  off ($S_1=off$ and $S_2=off$) and the current pass through the $R_\text{off}$ $-R_\text{bias}$ branch. 
The on state is implemented when both  switches are on ($S_1=on$ and $S_2=on$) and the equivalent resistor is $R_\text{off} // R_\text{on}$ is approximated by $R_\text{on}$ . The snapback state occurs when ($S_1=on$ and $S_2=off$) and the equivalent resistor is  $(R_\text{off} // R_\text{on}) -R_\text{bias}$, which is approximated by $-R_\text{bias}$. It should be noticed that, the snapback state has a negative slope originated from the negative resistance state represented by $-R_\text{bias}$.
Our goal is to translate these findings into a mathematical description that
can be implemented in circuit simulators in a cost efficient way.
This model consists of the topological connectivity, and compact models, of the elements that form the circuit, including extraneous components like parasitic resistance, capacitance,  and inductance. The mathematical model of the circuit is given by a system of nonlinear, coupled, differential algebraic equations (DAEs) as follows
\begin{equation}
F(x, x^{.}, t) = 0
\end{equation}
where $F(\cdot)$ represents Kirchhoff's current law (KCL) and $x$ is the state variable (or voltage drop). 
The dynamics of these state variables can be defined as follows\cite{7154394}
 \begin{equation} 
 i=\displaystyle f(v)+ \frac {d}{dt}q(v)
 \label{eq:simu}
 \end{equation} 
where $f$ represents the currents flowing in static branches of the equivalent network, $dq/dt$ represents the (capacitive) current flowing in time-dependent branches of the equivalent network. This formalism in \cite{7154394} already has been used to model an electrostatic discharge (ESD) snapback in \cite{6333317, 8509752}. It is worth to draw attention to the fact that OTS behavior is similar to that of the silicon controlled rectifier (SCR), however the SCR is a three terminals device \cite{chua1979nonlinear}. A new model for SCR device as two terminals has been described in \cite{chua1980device}. 
\subsection{Macro Model}
Based on the conceptual block diagram in Fig. \ref{fig:block}, the threshold switching device can be represented by two switches.

In order to to model the OTS and to implement the model in a circuit simulator by considering \eqref{eq:simu} in a compact and
calculation efficient way, an equivalent circuit has been proposed using only two bipolar junction transistors (BJTs)
and a self biased resistor to describe OTS characteristics
\cite{chua1983negative,chua1984negative}. 
The proposed 2T1R circuit exhibits NDR
characteristics,   which is suitable for modelling locally active memristors or threshold switching devices.
Thus, a two-terminal device with a 
NDR 
has been implemented in LTSPICE as shown in Fig. \ref{fig:circuit}.
The first transistor is a PNP type, the voltage drop across the base-collector junction is described as follows  
\begin{equation}
    v_{{BC}_{1}}= v_2 =  -v_R
    \label{BC_PNP}
\end{equation}
and the voltage drop across the emitter-base junction as
\begin{equation}
v_{\text{EB}_1}= v_1
\end{equation}
The second transistor is a NPN type, the voltage drop across the base-collector junction is defined as follows  
\begin{equation}
    v_{\text{BC}_2}= -v_2 = v_R
    \label{v2}
\end{equation}
and the voltage drop across the base-emitter junction as
\begin{equation}
v_{{\text{BE}}_2}= v_3
\end{equation}
It should be mentioned that, the pair of corresponding NPN and PNP transistors are complementary transistors with near identical characteristics to each other. This results to 
\begin{equation}
v_{1}= v_3
\label{v3}
\end{equation}
The voltage drop across the ideal OTS device ($Rs=0$ and $Rp= \infty$) can be written as 
\begin{equation}
v= v_1 +v_2 + v_3
\end{equation}
thus,
\begin{equation}
v-v_2 = v_1 + v_3
\end{equation}
Using \ref{v2} and \ref{v3}, the last equation can be rewritten as
\begin{equation}
v+v_R = 2v_1 
\end{equation}
or
\begin{equation}
\frac{v+v_R}{2} = v_3 = v_1 
\end{equation}
It should be noticed that, for the NPN transistor,
\begin{equation}
v_{\text{BE}_2}=\frac{v+v_R}{2}  
\end{equation}
On the other hand, for the PNP transistor,
\begin{equation}
v_{\text {EB}_1}=\frac{v+v_R}{2}=-v_{\text{BE}_1}
\label{BE_PNP}
\end{equation}
The phenomenon of an OTS device can be modeled mainly using  the BJT. From the transistor point of view, a simple model like the Ebers-Moll one fits our requirements for the circuit in
Fig. \ref{fig:circuit}\cite{ebers1954large}. 
The Ebers–Moll equation used to describe the emitter current of the PNP transistor in any operating region is given as follows
\begin{equation}
     i_{\text E_1}= I_s \cdot [e^{-v_{\text{BE}_1}/v_T} - e^{-v_{\text{BC}_1}/v_T} + \frac{1}{\beta_F} \cdot (e^{-v_{\text{BE}_1}/v_T} -1) ]
    \label{ebresIE}
\end{equation}
where
\\
$v_{\text{T}}$ is the thermal voltage $kT/q$ (approximately $26 mV$  at $300 K$ $\approx$ room temperature)
\\
$I_{\text{S}}$ is the reverse saturation current of the base–emitter diode (on the order of  \begin{math} 10^{-15}\end{math} 
to \begin{math}10^{-12}\end{math} amperes)
\\
${\beta_F}$ is the forward common emitter current gain (20 to 500)
\\
\\
replacing \eqref{BC_PNP} and \eqref{BE_PNP} in the last equation is given 
\begin{equation}
    i_{\text E_1}= I_s \cdot [e^{(v+v_{R})/2v_T} (1 + \frac{1}{\beta_F}) -e^{v_{R}/v_T} - \frac{1}{\beta_F}]
    \label{dc_final}
\end{equation}
For ideal OTS devices ($Rs=0$ and $Rp= \infty$) the total DC current flowing through the junctions \eqref{eq:simu} can be expressed as follows
\begin{equation}
    f(v)= i_{\text E_1} ~~.
\end{equation}
As mentioned above, the the emitter current of the PNP transistor is equal to the  emitter current of the NPN transistor ($i_{\text E_2}=i_{\text E_1}$).
The capacitive current flowing through the OTS device ($J_1$, $J_2$, and $J_3$ in Fig. \ref{fig:newcircuit}) can be modelled as follows  
\begin{equation}
    \frac{d}{dt}q(v)=  \frac{d}{dt}q(v_1) +  \frac{d}{dt}q(v_2) + \frac{d}{dt}q(v_3)
    \label{dq}
\end{equation}
The first term in the last equation is related to the charge between the emitter-base junction of the PNP transistor ($J_1$) and can be expressed as follows
\begin{equation}
    \frac{d}{dt}q(v_1)=  \frac{d [C_1\cdot v_1]}{dt}
    \label{jd}
    \end{equation}
        assuming the capacitance $C_1$ consists of diffusion capacitance  $C_{d_1}$ and junction capacitance  $C_{j_1}$, 
        the right hand side of \eqref{jd}
        can be written as
    \begin{equation}
   \frac{d}{dt}q(v_1)=   \frac{d(C_{d_1}+C_{j_1})}{dt} \cdot (\frac{v+v_2}{2}) + (C_{d_1}+C_{j_1}) \cdot \frac{d(\frac{v+v_2}{2})}{dt}
\end{equation}
where 
\begin{equation}
    v_1 = \frac{v+v_2}{2}
\end{equation}
and
\begin{equation}
    v_1 = v_3 .
\end{equation}
The third term in \eqref{dq} is related to the charge between the emitter-base junction of the NPN transistor ($J_3$). Since the pair of transistors are complementary, this term can be modelled as
\begin{equation}
    \frac{d}{dt}q(v_3)=  \frac{d}{dt}q(v_1)
    \end{equation}
These two terms are negligible in comparison with one related to the base-collector junctions of both transistors ($J_2$). The main capacitive current flowing through the OTS device can be found by applying  KCL as follows        
\begin{equation}
    \frac{d}{dt}q(v_2)= i_{C_{1}} -i_{B_{2}} - i_{Rb}- [\frac{d}{dt}q(v_1)+\frac{d}{dt}q(v_3)]
    \label{icap}
\end{equation}
The Ebers–Moll equation used to describe the collector current of the PNP transistor in any operating region is given as follows
\begin{equation}
   i_{C_{1}}= I_s \cdot [e^{-v_{\text{BE}_1}/v_T} - e^{-v_{\text{BC}_1}/v_T} - \frac{1}{\beta_R} \cdot (e^{-v_{\text{BC}_1}/v_T}-1) ]
    \label{ebresIC}
\end{equation}
where $\beta_R$ is the reverse common emitter current gain (0 to 20).
The Ebers–Moll equation used to describe the base current of the NPN transistor in any operating region is given as follows
\begin{equation}
    i_{B_{2}}= I_s \cdot [\frac{1}{\beta_F} \cdot (e^{v_{\text{BE}_2}/v_T} -1)+ \frac{1}{\beta_R} \cdot( e^{v_{\text{BC}_2}/v_T}-1)]
    \label{ebresIB}
\end{equation}
Using the Ebers–Moll equations mentioned above and substituting \eqref{icap} will yield
\begin{multline}
    \frac{d}{dt}q(v_2)= I_s \cdot [e^{(v+v_{R})/2v_T} (1 - \frac{1}{\beta_F}) - \\e^{v_{R}/v_T} \cdot(1+ \frac{2}{\beta_R}) +\frac{2}{\beta_R} + \frac{1}{\beta_F}] - \frac{v_R}{R_b} - \\ [\frac{d}{dt}q(v_1)+\frac{d}{dt}q(v_3)]
    \label{eq:Icapacitor_final}
\end{multline}
Finally, the current flow through the OTS device in \eqref{eq:simu} can be found by adding \eqref{dc_final} and \eqref{eq:Icapacitor_final}  as follows
 \begin{multline}
     i=  2\cdot I_{S}\cdot [ e^{ \frac{v+v_R}{2v_{T}}} - e^{v_R/v_{T}} \cdot ( 1+ \frac{1}{\beta_R}) + \frac{1}{\beta_R}] \ - \frac{v_R}{R_b}
     \label{eq:i}
 \end{multline}
 The macro model in Fig. \ref{fig:circuit} has been implemented in LTSPICE. Also, the derived equation \eqref{eq:i} from the macro model has been implemented in MATLAB.
 Fig. \ref{fig:a11} shows the $i\text{--}v$ curve for an physical OTS device fabricated by Western Digital Research, where 
$v_{th}$ is the turn on voltage, $I_{th}$ is the current at turn on event, $v_{offset}$ is the voltage across the OTS device when on, 
and $I_{hold}$ is the smallest current required to maintain the on state.

Fig. \ref{fig:b11} and Fig. \ref{fig:c11} show the $i\text{--}v$ curve of the implemented model in MATLAB extracted from the macro model, where  Fig. \ref{fig:d11} demonstrates the $i\text{--}v$ curve of the macro model implemented in LTSPICE. 
The behavior of our model in (\ref{eq:i}) has been inspected for different parameters as can be seen in Fig. \ref{fig:parameters}. 
For example,  Fig. \ref{fig:parameters1} shows $R_{ON}$ when considering a series resistor $R_{s}$=100 $\Omega$.
On the other hand Fig. \ref{fig:parameters2} demonstrates the $i\text{--}v$ curves for thermal voltage variations. However, for all other simulations, the thermal voltage $v_{\text{T}}$ has been set to $26 mV$ at $300 K$ room temperature. Moreover, the behavior of the model has been tested for the reverse saturation current variation as seen in Fig. \ref{fig:parameters3}. The assumption for the other cases is $I_{S}=10^{-14}A$. Furthermore, the behaviour of our model has been simulated for different bias resistor (snapback resistor) and plotted in Fig. \ref{fig:parameters4}. For the rest of the simulation result we assume $R_{b}=5 k\Omega$,  $R_{p}=100 k\Omega$, and  $R_{s}=200 \Omega$. It is worth drawing the reader's attention to the fact that the snapback makes the $i\text{--}v$ curve multi valued, so we keep track of the branch using the snapback state variable.
\subsection{Physical Model}
The OTS device can be represented by three junctions based on the macro model in Fig. \ref{fig:circuit},  $J_1$, $J_2$, and $J_3$.
The DC junction currents $f(v)$ in \eqref{eq:simu} are obtained by superposition of all currents.
Adopting the Ebers-Moll model with additional junction capacitances for the BJTs, and extending this model to the transient behavior, we get the OTS equivalent circuit in Fig. \ref{fig:newcircuit}. So, each
junction is represented by a diode, a capacitance, and
a current source in parallel.
The diode is equivalent to the DC characteristic of
the p-n junction current as follows
\begin{equation} 
  I_{J} =  I_{0}\cdot (e^{v_J/V_{T}} -1)
  \label{J}
  \end{equation} 
  where,
   \begin{equation} 
 I_{0} = \frac{I_s}{\alpha}
 \label{alph}
  \end{equation} 
  and
   \begin{equation}
        \alpha =\frac{\beta}{1+ \beta}
        \label{eq:albe}
  \end{equation}
The current source is defined as
\begin{equation} 
\alpha \cdot  I_{J}
\label{collect}
  \end{equation} 
The currents flowing in the static branch can be found by applying KCL as follows
\begin{equation} 
  f(v)= I_1 - \alpha_{R} \cdot  I_2
  \end{equation} 
  From \eqref{J} and \eqref{collect} the  last equation can be written as 
  \begin{equation} 
  f(v)=  I_{{0}_1}\cdot (e^{v_1/V_{T}} -1) - \alpha_{R} \cdot  I_{{0}_2}\cdot (e^{v_2/V_{T}} -1)
  \label{rep}
  \end{equation}
  replacing \eqref{alph} in \eqref{rep} yielding to  
   \begin{equation} 
  f(v)=  \frac{I_{s} }{\alpha_F}\cdot (e^{v_1/V_{T}} -1) - I_{s} \cdot   (e^{v_2/V_{T}} -1)
  \end{equation}
    hence,
  \begin{equation}
     f(v)= I_s \cdot [e^{v_{1}/v_T} - e^{-v_{2}/v_T} + \frac{1}{\beta_F} \cdot e^{v_{1}/v_T} ]
\end{equation}
or
\begin{equation}
     f(v)= I_s \cdot [e^{v_{1}/v_T} (1 + \frac{1}{\beta_F}) -e^{-v_{2}/v_T} - \frac{1}{\beta_F}]
\label{dc_final1}
\end{equation}
The capacitance $C_{total}$ in \eqref{eq:simu} consists of a junction capacitance
and a diffusion capacitance for each P-N junction (depicted in Fig. \ref{fig:newcircuit}) as follows
\begin{equation}
    C_{total}= C_j + C_d .
    \label{capacitane}
\end{equation}
Both capacitances are voltage dependent. We treat these junctions as Schottky junctions with the capacitance defined as follows \cite{6333317}
\begin{equation}
    C_{j}=\frac{C_{j0}}{(1-v_c/v_{j})^M},
\end{equation}
where $C_{j0}$ is the zero-bias capacitance, $v_{j}$ is the built-in potential, and $M$ is the grading coefficient.
The diffusion capacitance represents the minority
carrier charge. The different diffusion capacitances
must be summed as follows
\begin{equation}
    C_{d} \approx \Sigma \tau_{j} \frac{I_j}{V_T}
\end{equation}
The current flowing in time-dependent branch related to $v_2$ and $J_2$ is the main branch, and can be expressed as
\begin{equation}
    \frac{dq(v)}{dt} \approx \frac{dq(v_2)}{dt}
    \label{eq:apo}
\end{equation}
In order to find this current, we have applied KCL at $J_2$ as follows
\begin{multline}
\frac{dq(v_2)}{dt} = \alpha_{F} \cdot  I_1 - I_2 - I_3 +\\ \alpha_{R} \cdot  I_2 + \alpha_{F} \cdot  I_3 - I_2 -\frac{v_R}{R_b}
\end{multline}
replacing \eqref{J} and \eqref{collect} in the last equation gives
\begin{multline}
\frac{dq(v_2)}{dt} = \alpha_{F} \cdot  I_{{0}_1}\cdot (e^{v_1/V_{T}} -1)  - 2 \cdot I_{{0}_2}\cdot (e^{v_2/V_{T}} -1) \\ -   I_{{0}_3}\cdot (e^{v_3/V_{T}} -1) + \alpha_{R} \cdot   I_{{0}_2}\cdot (e^{v_2/V_{T}} -1) +\\ \alpha_{F} \cdot  I_{{0}_3}\cdot (e^{v_3/V_{T}} -1) -\frac{v_R}{R_b}
\end{multline}
where the $\alpha_\_$ are the scaling factors of the current sources in Fig. \ref{fig:newcircuit}.
Using \eqref{alph} results to
\begin{multline}
\frac{dq(v_2)}{dt} = I_{s}\cdot (e^{v_1/V_{T}} -1)  - 2 \cdot \frac{I_{s}}{\alpha_R} \cdot (e^{v_2/V_{T}} -1) \\ -   \frac{I_{s}}{\alpha_F} \cdot (e^{v_3/V_{T}} -1) + I_{s} \cdot (e^{v_2/V_{T}} -1) +\\ I_{s} \cdot (e^{v_3/V_{T}} -1) -\frac{v_R}{R_b} \end{multline}
hence,
\begin{multline}
\frac{dq(v_2)}{dt} = I_{s}\cdot [ 2 \cdot (e^{v_1/V_{T}} -1)  + (e^{v_2/V_{T}} -1) \cdot ( 1- \frac{2}{\alpha_R} ) \\ -   \frac{1}{\alpha_F} \cdot (e^{v_1/V_{T}} -1) -\frac{v_R}{R_b}
\end{multline}
replacing \eqref{eq:albe} in the last equation yields 
\begin{multline}
    \frac{d}{dt}q(v_2)= I_s \cdot [e^{v_1/2v_T} (1 - \frac{1}{\beta_F}) - \\e^{v_{R}/v_T} \cdot(1+ \frac{2}{\beta_R}) +\frac{2}{\beta_R} + \frac{1}{\beta_F}] - \frac{v_R}{R_b} 
    \label{Icapacitor_final1}
\end{multline}
Finally, the current flow through the OTS device in \eqref{eq:simu} can be found by taking the sum of \eqref{dc_final1} and \eqref{Icapacitor_final1}  as follows
 \begin{multline}
     i=  2\cdot I_{S}\cdot [ e^{ \frac{v_1}{v_{T}}} - e^{v_R/v_{T}} \cdot ( 1+ \frac{1}{\beta_R}) + \frac{1}{\beta_R}] \ - \frac{v_R}{R_b}
     \label{eq:i12}
 \end{multline}
\begin{table}[]
\renewcommand{\arraystretch}{1.3}
\centering
\caption{The used parameters to fit data in Fig. \ref{fig:fitting} and their corresponding values.}
\label{tab:parameter}
\begin{tabular}{|c|c|}
\hline
\textbf{Parameter} & \textbf{Value} \\ \hline
      $I_s$    &     $10^{-14} A$      \\ \hline
      $\beta_F$    &   $250$        \\ \hline
      $V_T$    &     $0.0259V$      \\ \hline
       $K$   &   $0.7$        \\ \hline
       $I_{State}$   & $1 \mu A$          \\ \hline
      $R$    &    $1M \Omega$       \\ \hline
       $C$   &     $1nF$      \\ \hline
       $v_{th}$   &   $2.4V$        \\ \hline
       $i_{th}$   &    $1 \mu A$        \\ \hline
         $R_{b}$   &     $5 k \Omega$         \\ \hline
\end{tabular}
\end{table}
\subsection{Internal State Variable}
 Based on the physical model, we have implemented a delay model to switch the state between its off and on values, 0 and 1, respectively, as seen in Fig. \ref{fig:a1}. It should be noticed that the physical model allowed constructing the delay model using few elements relative to the second junction in the physical model. For instance, the capacitor $C_2$ is proportional to all the parallel capacitors in the second junction $J_2$, the resistor $R_2$ is proportional to the bias resistor and the diodes in parallel, and the current source $I_{State}$ is linked to the current sources.

The value of the current source driving the delay model in Fig. \ref{fig:a1} is set to $1 \mu A$ when the voltage across the device exceeds or reaches the threshold voltage, $v_{th}$ (where $i_{th}$ is the corresponding current) . Otherwise, the value of the current source is zero when the voltage across the device drops below the threshold voltage. Applying KCL, the delay model can be expressed as follows
\begin{equation}
    I_{State}= \frac{v_{R}}{R_2} + C_2 \cdot \frac{dv_R}{dt} 
\end{equation}
We introduce an internal state variable $State$ or $  \zeta  \in [0,1]$, the value of the which represents a voltage that can be found from the solution of the previous equation as follows
\begin{equation}
    \zeta(t) =  I_{State} \cdot R_2 - e^{\frac{-t}{R_{2}C_2}} 
    \label{State}
\end{equation}
The values of the internal state variable are plotted  in Fig. \ref{fig:state}. Let us consider a scaling factor $K$, the internal voltage drop across the bias resistor can be written as follows  
\begin{equation}
  v_R=  K \cdot  \zeta 
  \label{KState}
\end{equation}
The internal voltage drop across the bias resistor has been plotted in Fig. \ref{fig: Internal State Variable} for different $K$ values. 
As a result, the on state voltage can be simply characterized by the voltage drop across the bias resistor as follows
\begin{multline}
 v_{on}= {ln(\beta_F/(\beta_F+1))^{2 \cdot V_T}} -v_R+ \\
 {ln((i_{on}/Is)^{2 \cdot V_T}} +(1/\beta_F)+exp(v_R/V_T)) 
 \label{model1}
\end{multline}
or the on state current can be written as 
\begin{multline}
     i_{on}=  2\cdot I_{S}\cdot [ e^{ \frac{v_1}{v_{T}}} - e^{v_R/v_{T}} \cdot ( 1+ \frac{1}{\beta_R}) + \frac{1}{\beta_R}] \ - \frac{v_R}{R_b}
\end{multline}
 The off state current is essentially a leakage current, and it is modeled by a modified Shockley diode equation as
 \begin{equation}
  i_\text{off} = 10 ^{  \log_{10}[i_{th}] - \Delta \cdot (v-v_{th}) } -i_{0}
   \label{model2}
\end{equation}
where,
 \begin{equation}
\Delta= \log_{10}[i_{th}] \cdot (\frac{1}{v_{th}} )
\end{equation}
and
\begin{equation}
  i_{0} = 10 ^{  \log_{10}[i_{th}] + \Delta \cdot v_{th}}
\end{equation}
With the state variable $\zeta$, $S=0/1$ indicating off/on states, both equations for the current $i$ using \eqref{eq:simu} can be combined into one formula as follows  
\begin{equation}
  i= C \cdot \frac {d}{dt}v_R + (1- S) \cdot i_\text{off}(v)+ S \cdot i_\text{on}(v)
  \label{eq:final model}
  \end{equation} 
   It should be noticed that a linear capacitor has been assumed as an approximation $C$ $\approx$ $C_2$, as we concentrate mainly on the modelling of $i\text{--}v$ snapback in this article. A more accurate charge model explained in \eqref{capacitane} can be used in this formulation without modifying the equation structure. Fig. \ref{fig:Q} shows the derivative of the internal voltage drop across the bias resistor and the charge model for two different values of capacitance.  
  
  The $i\text{--}v$ characteristics using the compact model of \eqref{eq:final model} are shown in Fig. \ref{fig:i-v simulation}. 
  We have validated the model with experimental data for a physical OTS device which consisted of a Se-based OTS film approximately 15nm thick with carbon electrodes. The OTS layer and electrodes were patterned into a pillar of approximately 40nm diameter.  Data was collected by applying a voltage pulse, with an on-chip resistor used to limit the current after OTS thresholding.
The parameters values for this fitting are listed in Table \ref{tab:parameter}. By overlaying the fitting line with the measurement, we show that the model reproduces the $i\text{--}v$ characteristics well, not just in the on and off states, but also in the snapback region as can be seen in Fig. \ref{fig:fitting}. 
\section{Conclusion}
\label{sec:discussion}
In this paper, we have presented a new compact model for efficient circuit-level simulations of threshold switching devices. As a first step, a macro model has been implemented in LTSPICE. Based on this macro model, a descriptive model has been extracted and implemented in MATLAB. The macro model has been extended into a physical model to capture the switching process by adding a delay through additional components connected to the second junction $J_2$ in the physical model. This delay model introduces an internal state variable, which is necessary to convert the descriptive model to a compact model and to parameterize it in terms of easily extractable electrical parameters that represent device behaviour. Finally, we verified our model by fitting $i\text{--}v$ measured data of the physical OTS devices. 
\section*{Acknowledgment}
This work has been funded by Western Digital Corporation, California, U.S.A.

\end{document}